\documentclass[aps,prl,twocolumn,superscriptaddress,longbibliography,nobibnotes]{revtex4-2}
\usepackage{amsmath}
\usepackage{amssymb}
\usepackage{mathrsfs}
\usepackage{dsfont}
\usepackage[pdftex,colorlinks=true,urlcolor=blue,linkcolor=blue,citecolor=blue,breaklinks=true,bookmarks=false]{hyperref}
\usepackage{graphicx}
\usepackage{float}
\newcommand{\push}{\hspace{0.05cm}}
\newcommand{\pull}{\hspace{-0.05cm}}
\newcommand\sbullet[1][.5]{\mathbin{\vcenter{\hbox{\scalebox{#1}{$\bullet$}}}}}

    \usepackage{multibib} %right-align reference numbers

    \usepackage{titlesec}
    \titleformat*{\section}{\large \bfseries}
    \titleformat*{\subsection}{\bfseries}
    \titleformat*{\subsubsection}{\itshape}

\begin{document}

\title{Quantum walk of two anyons across a statistical boundary}

\author{Liam L.H. Lau}
\email[E-mail: ]{liam.lh.lau@physics.rutgers.edu}
\affiliation{Gonville \& Caius College, University of Cambridge, Trinity Street, Cambridge, CB2 1TA, United Kingdom\looseness=-1}
\author{Shovan Dutta}
%\thanks{Present Address: Max Planck Institute for the Physics of Complex Systems}
\email[E-mail: ]{sdutta@pks.mpg.de}
\affiliation{T.C.M. Group, Cavendish Laboratory, University of Cambridge, JJ Thomson Avenue, Cambridge CB3 0HE, United Kingdom\looseness=-1}

\date{\today}

\begin{abstract}
We model a quantum walk of identical particles that can change their exchange statistics by hopping across a domain wall in a 1D lattice. Such a ``statistical boundary'' is transparent to single particles and affects the dynamics only by swapping multiple particles arriving together. We find that the two-particle interference is dramatically altered by reflections of these bunched waves at the interface, producing strong measurable asymmetries. Depending on the phases on the two sides, a bunched wavepacket can get completely reflected or split into a superposition of a reflected wave and an antibunched wave. This leads to striking dynamics with two domain walls, where bunched waves can get trapped in between or fragment into multiple correlated single-particle wavepackets. These findings can be realized with density-dependent hopping in present-day atomic setups and open up a new paradigm of intrinsically many-body phenomena at statistical boundaries.
\end{abstract}

\maketitle

{\it Introduction.---}The behavior of identical quantum particles is dictated by their exchange statistics, i.e., the phase ($\theta$) acquired by the wavefunction when two particles are exchanged \cite{Cohen2010quantum}. Bosons, with $\theta=0$, lead to blackbody radiation and Bose-Einstein condensates, whereas fermions, with $\theta=\pi$, form neutron stars and the periodic table of elements. More exotic particles, with $0 < \theta < \pi$, can exist only in low dimensions \cite{Leinaas1977, Wilczek1982, Canright1990}. These ``anyons'' are found as surface excitations on a fractional quantum Hall state \cite{Stern2008} or spin liquids \cite{Yao2007} and have been observed in recent experiments \cite{Bartolomei2020, Nakamura2020}, but they can also arise in one dimension (1D) \cite{Haldane1991}. Their fractional statistics has fueled intense research \cite{wilczek1990fractional, Kitaev2006, Greschner2018} and is the basis for topological quantum computing protocols \cite{Nayak2008, Lahtinen2017}.

The rise of controllable atomic and photonic platforms has meant one can engineer particle statistics in experiments \cite{Georgescu2014, Altman2020quantum}. In particular, Keilmann et al. \cite{Keilmann2011} have shown anyons on a 1D lattice are equivalent to bosons with density-dependent hopping which has been realized in atomic setups \cite{Juergensen2014, Meinert2016, Clark2018, Goerg2019, Schweizer2019}. Subsequently, Greschner and Santos \cite{Greschner2015} showed the statistics of these anyons is fully tunable by a Raman laser. Motivated by such possibilities, theories have found rich ground states \cite{ArcilaForero2016, Lange2017, Tang2015} and dynamics \cite{Hao2012, Wang2014, Piroli2017}. In these studies, the exchange phase, set by complex hopping amplitudes, is spatially uniform. Yet, the protocol in Ref.~\cite{Greschner2015} can be extended to nonuniform phases, which produces an intriguing scenario where anyons change their statistics by simply hopping across a domain wall. Here we explore new physics resulting from such a ``statistical boundary'' by modeling two-particle walks that can be monitored in experiments.

Two-body walks give a clean signature of the underlying statistics through the interference of multiple two-particle pathways, which produces bunching of bosons and antibunching of fermions, exemplified by the %Hanbury Brown-Twiss \cite{Brown1956} and 
Hong-Ou-Mandel \cite{Hong1987, Kaufman2018} effect. Such walks have been realized in atom traps \cite{Jeltes2007, Schellekens2005, Preiss2015} and photonic circuits \cite{Peruzzo2010, Sansoni2012, Matthews2013}, and have key applications in quantum computing \cite{Childs2009, VenegasAndraca2012}. Two-body interference of anyons with a given statistics has also been examined \cite{Wang2014, Sansoni2012, Matthews2013, Campagnano2012}. In 1D models \cite{Keilmann2011}, the anyons can have double occupancies even at $\theta=\pi$ to allow for exchange, so in this limit they behave as ``pseudofermions'' that retain some bunching behavior \cite{Wang2014}.

\begin{figure}
\centering
\includegraphics[width=\columnwidth]{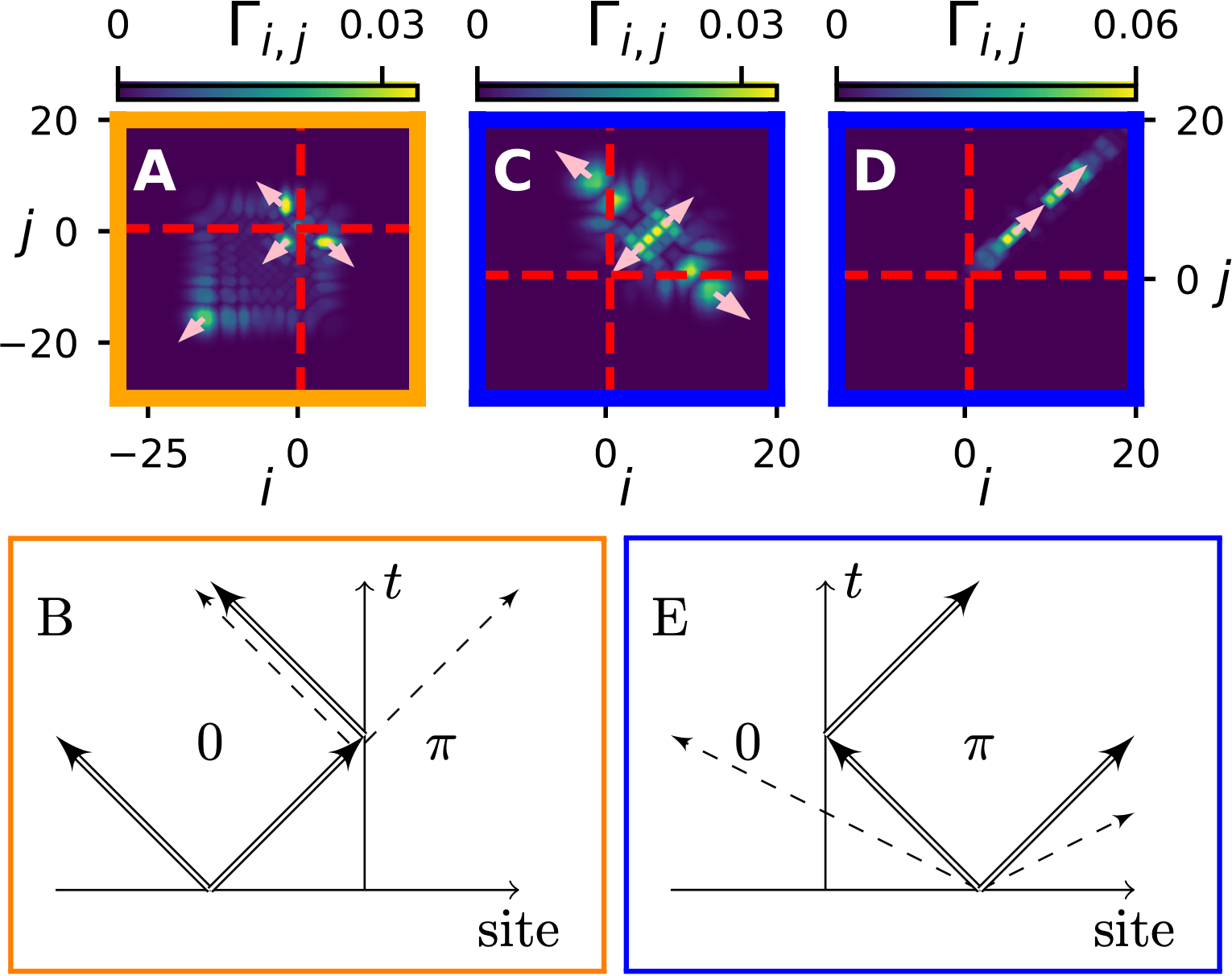}
\caption{\label{reflectionfig}Reflection of bunched waves off a 0-$\pi$ statistical interface after two anyons with $U=0$ are released from (A, B) the boson side, $j=-6,-5$, and (C, D, E) the pseudofermion side, $j=5,6$. (A) Two-body correlations $\Gamma_{i,j}$ at $Jt = 5.9$, showing a bunched wave being split into a reflected bunched wave and an antibunched wave. (C) $\Gamma_{i,j}$ at $Jt = 4$, showing fast antibunched waves passing through the boundary and a slow bunched wave arriving at the interface. (D) $\Gamma_{i,j}$ at $Jt = 47.3$, showing the bunched wave is completely reflected. (B, E) Sketch of the dynamics: double (single) arrows show bunched waves (single particles).
}
\end{figure}

\begin{figure}[b]
\centering
\includegraphics[width=\columnwidth]{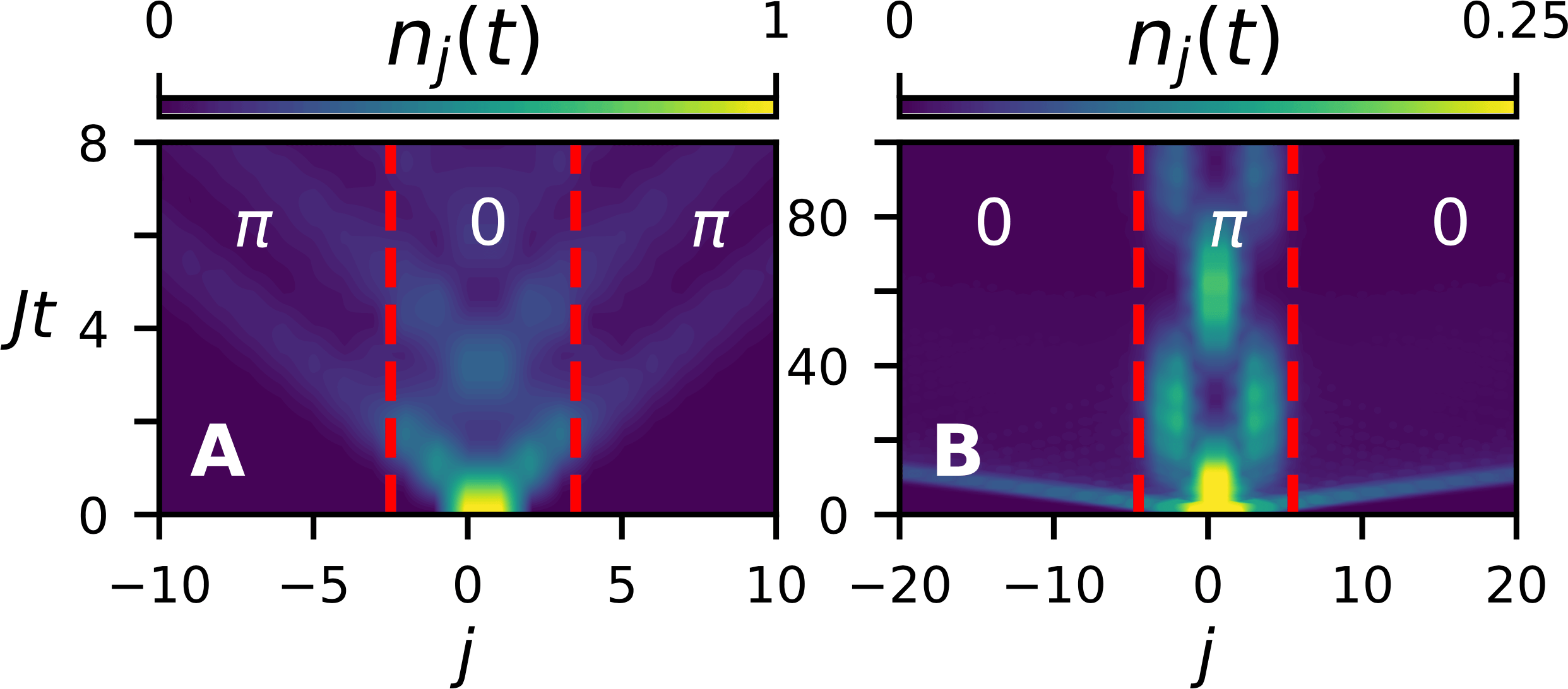}
\caption{\label{wellfig}Back-and-forth reflection of bunched waves inside a statistical well, after release of two anyons at $j=0,1$ with $U=0$. (A) $\pi$-0-$\pi$ interface: $\theta_j = 0$ for $-2 \leq j \leq 3$. Reflections are lossy, as in Fig.~\ref{reflectionfig}B, producing fragmented single-particle waves. (B) 0-$\pi$-0 interface: $\theta_j = \pi$ for $-4 \leq j \leq 5$. Slow bunched waves undergo lossless reflections, and a fast antibunched wavefront flies off at short times, as in Fig.~\ref{reflectionfig}E.
}
\end{figure}

We consider a statistical boundary where the exchange phase is $\alpha$ on one side and $\beta$ on the other side. On either side, the propagation occurs in the form of bunched and antibunched waves. Antibunched waves are unaffected by the interface since the two particles move separately. On the other hand, bunched waves are strongly transformed: For a boson-pseudofermion ($0$-$\pi$) interface, we show that a bunched wave incident from the fermionic side is completely reflected, whereas one from the bosonic side is coherently split into a reflected bunched wave and an antibunched wave (Fig.~\ref{reflectionfig}). Hence, the long-time dynamics are very sensitive to initial conditions, changing dramatically as one crosses the boundary. We fully characterize this physics by global number asymmetries that can be measured, e.g., with a quantum gas microscope \cite{Preiss2015}. We predict striking consequences, including a statistical well that can trap or successively fragment bunched particles (Fig.~\ref{wellfig}). These features are most prominent at weak onsite interactions and large phase jumps at the boundary, which can both be tuned by Raman lasers \cite{Greschner2015}. %Our study opens up investigations of correlated many-body dynamics at statistical domain walls.

{\it Model.---}Anyons on a 1D lattice with a given exchange phase $\theta$ are defined by the commutation relations $\hat{a}_i \hat{a}_j = e^{{\rm i}\theta} \hat{a}_j \hat{a}_i $ and $\smash{\hat{a}_i \hat{a}_j^{\dagger} = e^{-{\rm i}\theta} \hat{a}_j^{\dagger} \hat{a}_i} $ for all $i<j$, where $\smash{\hat{a}_i^{\dagger}}$ creates an anyon at site $i$. There is some freedom in choosing the on-site statistics ($i=j$) \cite{Girardeau2006}. We follow the convention in Refs.~\cite{Keilmann2011, Kundu1999, Bonkhoff2020bosonic} where this is bosonic; i.e., $\smash{[\hat{a}_i,\hat{a}_i^{\dagger}] = 1}$, so multiple anyons can occupy the same site and exchange positions. This choice is further motivated by a Jordan-Wigner transform that maps such anyons to interacting bosons which can be studied experimentally \cite{Keilmann2011}.

We introduce anyons with a spatially varying exchange phase through a modified Jordan-Wigner (JW) map,
\begin{equation}
\hat{a}_j := e^{{\rm i}\sum_{k<j} \theta_k \hat{n}_k} \hat{b}_j \;,
\label{anyondef}
\end{equation}
where $\hat{b}_j$ are the boson operators and $\smash{\hat{n}_k := \hat{b}_k^{\dagger} \hat{b}_k = \hat{a}_k^{\dagger} \hat{a}_k}$ is the occupation at site $k$. Using the relations $\smash{[\hat{b}_i, \hat{b}_j] = 0}$ and $\smash{[\hat{b}_i, \hat{b}_j^{\dagger}] = \delta_{ij}}$, we find the anyonic commutations
\begin{equation}
\hat{a}_i \hat{a}_j = e^{{\rm i}\theta_i} \hat{a}_j \hat{a}_i  \quad \text{and} \quad
\hat{a}_i \hat{a}_j^{\dagger} = e^{-{\rm i}\theta_i} \hat{a}_j^{\dagger} \hat{a}_i
\label{anyoncomms}
\end{equation}
for $i<j$; i.e., the exchange phase is set by the ``left'' site. This loss of reflection symmetry arises from Eq.~\eqref{anyondef} and is characteristic of anyons \cite{wilczek1990fractional}. We are interested in cases where all exchanges are local and $\theta_k$ varies sharply across a domain wall, so particles on either side of the wall have well-defined and distinct statistics. Such an interface is ``invisible'' to single particles and affects the physics only by exchanging anyons between the two sides.

\begin{figure}
\centering
\includegraphics[width=\columnwidth]{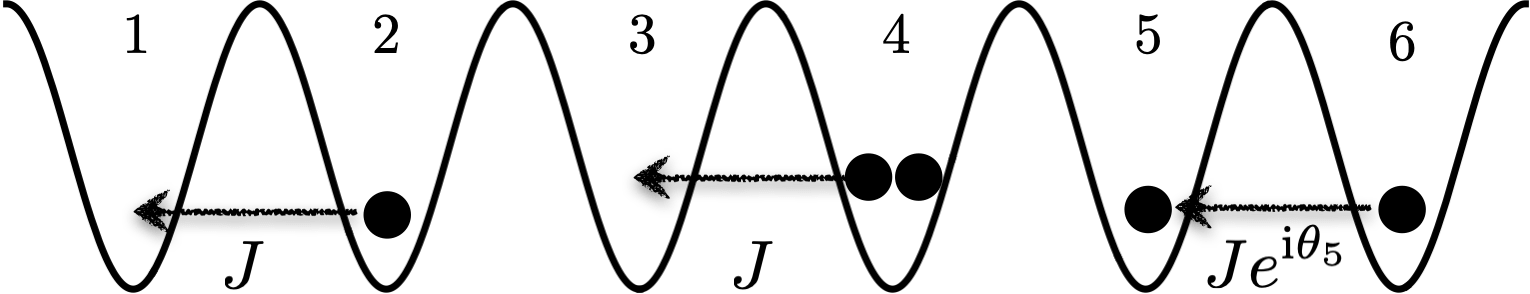}
\caption{\label{tunnelingfig}Schematic of density-dependent tunneling in Eq.~\eqref{hamilboson}.}
\end{figure}

To probe the resulting dynamics, we adopt the Anyon-Hubbard Hamiltonian \cite{Keilmann2011}
\begin{equation}
\hat{H} = 
-J \sum\nolimits_j \big(\hat{a}_j^{\dagger} \hat{a}_{j+1} + \text{H.c.}\big) 
+ U \sum\nolimits_j \hat{n}_j (\hat{n}_j -1)/2 \push,
\label{hamilanyon}
\end{equation}
where $J$ is the nearest-neighbor tunneling and $U$ is an on-site interaction. Crucially, this Hamiltonian maps onto a Hubbard model for bosons via Eq.~\eqref{anyondef},
\begin{equation}
\hat{H} = 
-J \sum_j \big(\hat{b}_j^{\dagger} \hat{b}_{j+1} e^{{\rm i} \theta_j \hat{n}_j} + \text{H.c.}\big) 
+ \frac{U}{2} \sum_j \hat{n}_j (\hat{n}_j -1) \push.
\label{hamilboson}
\end{equation}
Here $\theta_j$ gives an occupation-dependent Peierls phase; i.e., a hop from site $j+1$ to $j$ yields an additional phase depending on the occupation of site $j$, as shown in Fig.~\ref{tunnelingfig}. Such phases have been realized in shaken optical lattices \cite{Meinert2016, Clark2018, Goerg2019, Schweizer2019} in a quest to simulate dynamical gauge fields \cite{Banuls2020}. Furthermore, theoretical studies have shown that Peierls phases of the specific form in Eq.~\eqref{hamilboson} can be engineered by Raman-assisted tunneling \cite{Keilmann2011, Greschner2015} and lattice shaking \cite{Straeter2016}. Of these, the protocol in Ref.~\cite{Greschner2015} is particularly flexible and readily generalized to nonuniform $\theta_j$ by spatially modulating a Raman laser. Note that $\theta_j$ mediates an effective interaction between the {\it bosons}, which is distinct from the onsite interaction $U$. This is a consequence of the fractional anyonic statistics.

%We study two-body walks with a statistical boundary. 
At any time $t$, the two-particle state can be expressed in terms of the boson operators as
\begin{equation}
|\Psi(t)\rangle = 
\sum\nolimits_j \pull d_j(t) \push \hat{b}_j^{\dagger 2} |0\rangle / \sqrt{2} \push
+ \sum\nolimits_{i<j} \pull c_{i,j}(t) \push \hat{b}_i^{\dagger} \hat{b}_j^{\dagger} |0\rangle \push,
\end{equation}
where $|0\rangle$ is the vacuum and $\smash{\sum_j |d_j|^2 + \sum_{i<j} |c_{i,j}|^2 = 1}$ for normalization. Bunching or antibunching of the particles shows up in the density-density correlations $\Gamma_{i,j} :=  \langle \hat{n}_i \hat{n}_j \rangle - \delta_{ij} n_i  =   2 |d_j|^2 \delta_{ij} + |c_{i,j}|^2 (1 - \delta_{ij})$, and can be measured experimentally \cite{Schellekens2005, Preiss2015, Peruzzo2010, Sansoni2012, Matthews2013}. Here $n_i := \langle \hat{n}_i \rangle = \sum_{j} \Gamma_{i,j}$. We focus on neighboring initial states where the anyons most strongly influence one another. We find the coefficients $d_j$ and $c_{i,j}$ by exact diagonalization, using a large grid to avoid reflection from edges.

\begin{figure}
\centering
\includegraphics[width=\columnwidth]{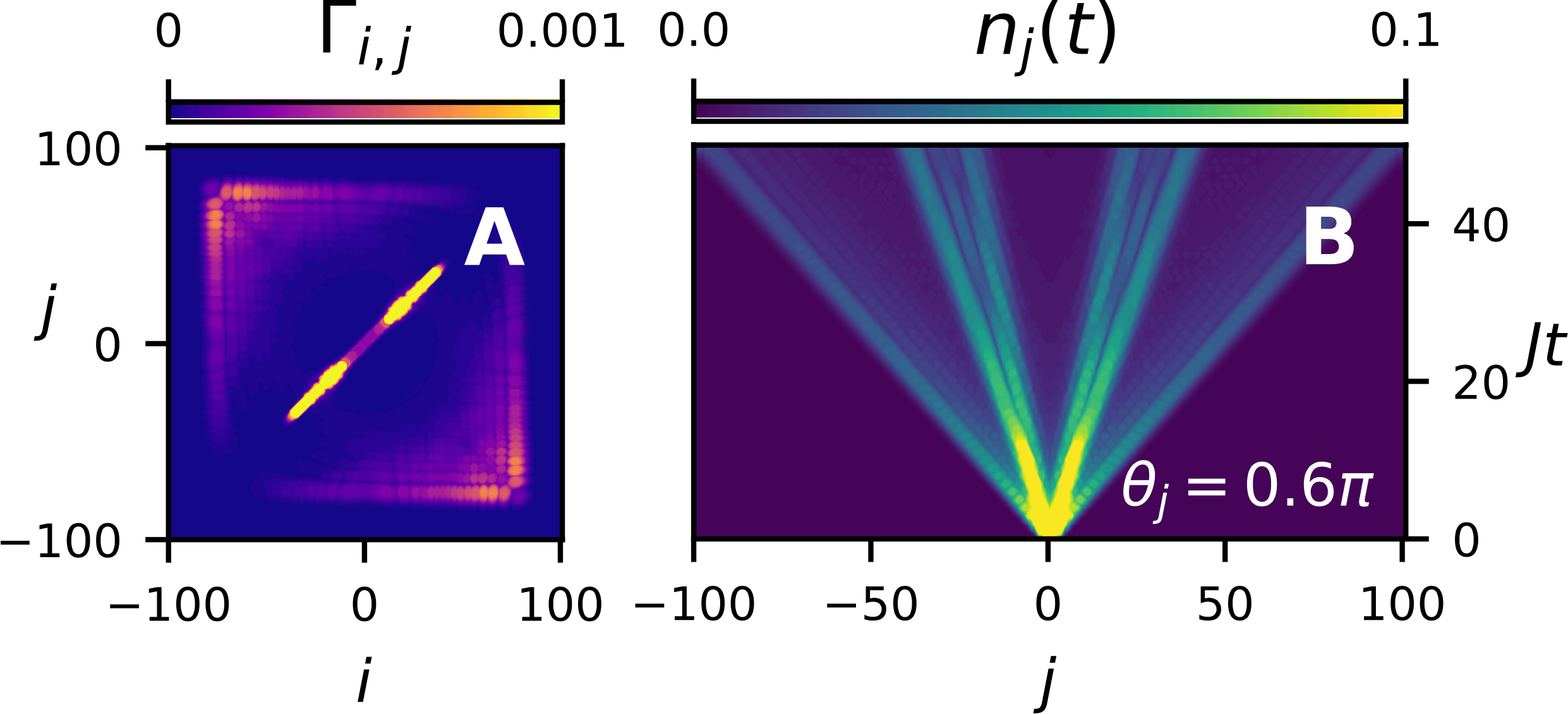}
\caption{\label{uniformfig}Motion of two anyons with exchange phase $\theta_j = 0.6\pi$ and onsite interaction $U=0$, starting from sites $j=0,1$. (A) Density-density correlations $\Gamma_{i,j}$ at $Jt = 40$ and (B) density $n_j(t)$, showing fast antibunched and slow bunched waves.}
\end{figure}

{\it Uniform case.---}Before considering a domain wall, we discuss the physics in the uniform case \cite{Wang2014}, $\theta_j=\phi$, after the particles are released from adjacent sites. For $\phi=0$ and $U=0$, one has pure bosons that spread out in both directions as bunched waves with speed $2J$ (in units of lattice spacing, with $\hbar=1$). For $\phi\neq 0$, one instead finds a superposition of bunched and antibunched propagation, as shown in Fig.~\ref{uniformfig}. This can be seen either as a result of the anyonic statistics [Eq.~\eqref{anyoncomms}] or that of the occupation-dependent hopping of the JW bosons [Eq.~\eqref{hamilboson}]. The antibunched wave describes the two particles moving in opposite directions at speed $2J$, while the bunched wave is significantly slower. As $\phi$ is increased, the antibunching becomes more prominent and the bunched waves slow down further. However, even in the pseudofermion limit, $\phi=\pi$, the latter carry almost half the total weight at a speed $v_{\text{slow}} \approx J/5$ (see Supplement \cite{supplement}). One obtains stronger antibunching by increasing $U$. For $U/J \to \infty$, Eq.~\eqref{hamilboson} reduces to hard-core bosons that behave like free fermions \cite{Preiss2015} regardless of $\phi$. So the exchange phase is more relevant at smaller $U$.

The slow bunched wave can be explained qualitatively for $\phi \lesssim 1$ by calculating the scattering length \cite{Greschner2015}, which mimics an effective repulsion
\begin{equation}
U_{\text{eff}} = 4J \tan^2(\phi/2)
\label{Ueff}
\end{equation}
at $U=0$, leading to slow bound pairs \cite{Preiss2015}. However, this picture breaks down for large angles \cite{Kolezhuk2012}. In particular, at $\phi=\pi$, $U_{\text{eff}} \to \infty$, which predicts free-fermionic behavior and does not support bunching.

{\it Statistical boundary.---}We consider a sharp domain wall such that $\theta_j = \alpha$ for $j \leq 0$ and $\beta$ for $j>0$. We focus on a boson-pseudofermion interface, i.e., $\alpha=0$ and $\beta=\pi$, which produces the most striking departures. In a later section, we discuss how these effects fade gradually as one reduces the phase jump, increases $U$, or makes the interface less sharp. Recall that $U$ and $\theta_j$ are fully tunable by the protocol in Ref.~\cite{Greschner2015}.

\begin{figure}
\centering
\includegraphics[width=\columnwidth]{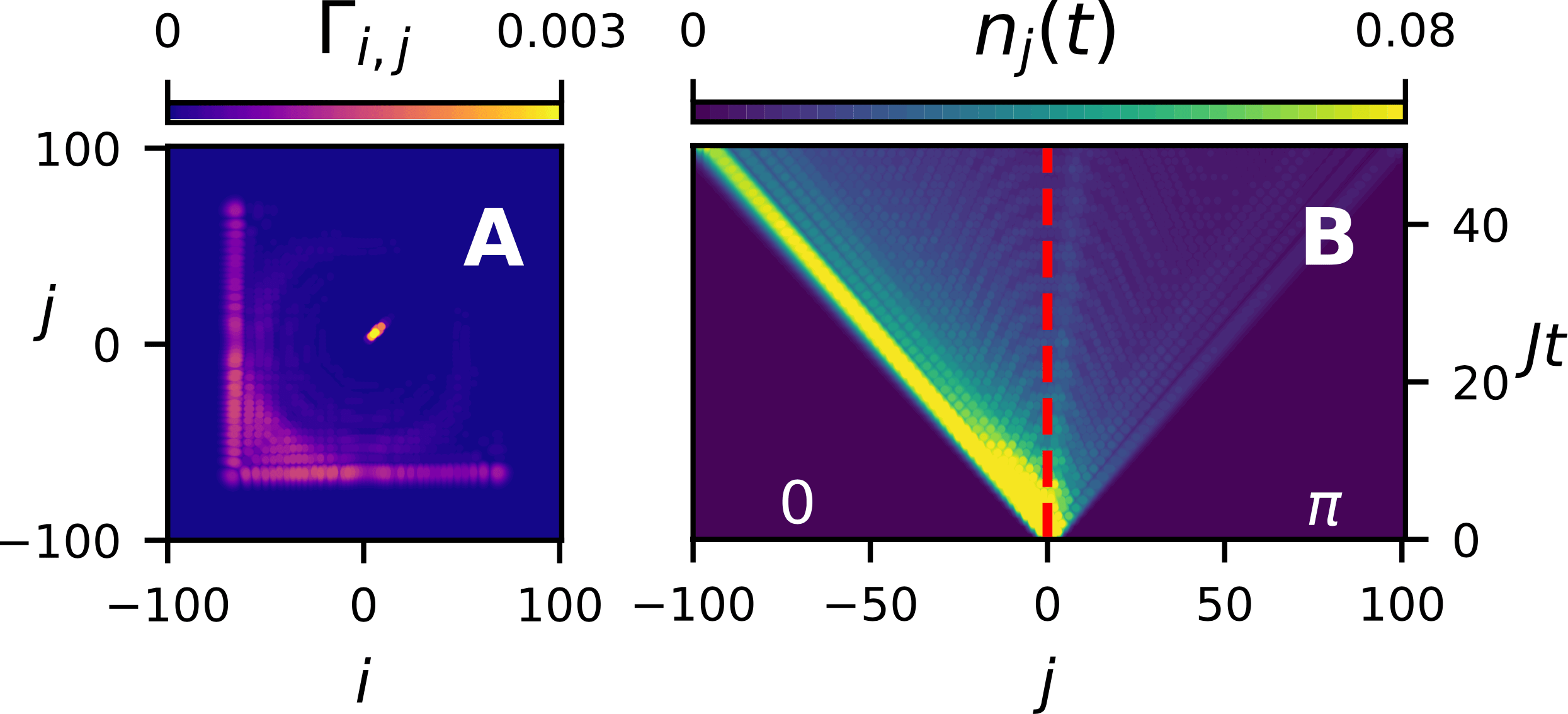}
\caption{\label{asymfig}Strongly asymmetric quantum walk two anyons with $U=0$ in the presence of a boson-pseudofermion interface: $\theta_j = 0$ for $j \leq 0$ and $\theta_j=\pi$  for $j>0$, after they are released from sites $j=0,1$. (A) Two-body correlations at $Jt = 35$ and (B) density profile as a function of time, showing a pronounced bunched wavefront in the bosonic side and a weak, slow bunched wave in the pseudofermion side.
}
\end{figure}

Figure~\ref{asymfig} shows what happens if the two particles are released from sites $j=0,1$, straddling the interface. An extremely skewed evolution ensues in which most of the weight flies off into the bosonic region as a bunched wave moving at speed $2J$. This is accompanied by some remnant antibunching. The remainder is barely visible as a weak bunched wave moving into the pseudofermion side at speed $v_{\text{slow}}$, carrying less than 3\% of the total weight. This is in stark contrast to the symmetric walk in Fig.~\ref{uniformfig}. Here the initial state can be considered ``bosonic,'' as the first hop does not yield any Peierls phase (see Fig.~\ref{tunnelingfig}), so we expect a bunched wave in the boson side. One might also anticipate less transmission to the pseudofermion region due to the effective repulsion [Eq.~\eqref{Ueff}]. However, as we pointed out before, this is only a qualitative picture. Below we show the strong asymmetry originates from a characteristic reflection of the bunched waves off the domain wall. %that is characteristic of such domain walls.

To deconstruct this effect of the interface, we consider initial states farther away from it, so that the incident waves are clearly discernible. Figure~\ref{reflectionfig}A shows an example where the anyons are released well inside the bosonic region at $j=-6,-5$. As in the uniform case, they start spreading out as bunched waves in both directions. When the right-moving front arrives at the interface, we find it is coherently split into two parts. One of these is reflected as a bunched wave and the other turns into antibunched motion, where one particle enters the pseudofermion side and the other goes back to the boson side. This process is sketched in Fig.~\ref{reflectionfig}B. Note that no bunched waves pass through the interface, which gives rise to the asymmetry in Fig.~\ref{asymfig}. The dynamics are even more striking when the anyons are released in the pseudofermion side. Here one has two different timescales as in Fig.~\ref{uniformfig}. There is a fast outward spreading where the anyons travel in opposite directions. Being solo, the left-moving anyon cannot see the interface and passes straight through, as in Fig.~\ref{reflectionfig}C. Much later, the slow bunched wave arrives and gets completely reflected, as shown in Figs.~\ref{reflectionfig}D and \ref{reflectionfig}E.

The reflection of bunched waves incident from the boson side can be approximated by using effective hard-core interactions for $j>0$, in accordance with Eq.~\eqref{Ueff}. However, this recipe fails to capture the dynamics for pseudofermionic initial states (see Supplement \cite{supplement}). The lack of transmission of the bunched waves is consistent with a large difference in group velocity between the two sides. However, we emphasize this is not single-particle physics, but a result of destructive interference between two-particle paths. 
%, but a consequence of anyonic statistics.

\begin{figure}
\centering
\includegraphics[width=\columnwidth]{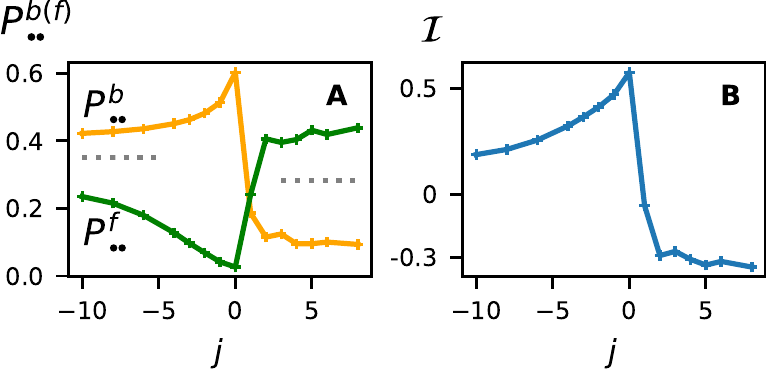}
\caption{\label{metricsfig}Long-time asymmetries ($Jt=100$) after the anyons are released from sites $j, j+1$, with the interface in Fig.~\ref{asymfig}. (A) Probability of finding both particles in left (orange) and right (green) halves. Dotted lines show probabilities for uniform $\theta_j$: $\smash{P^{b(f)}_{\sbullet[0.7]\sbullet[0.7]}} (0) \approx 0.35$ and $\smash{P^{b(f)}_{\sbullet[0.7]\sbullet[0.7]}} (\pi) \approx 0.28$. (B) Relative number imbalance between the two halves. The sudden drop at $j=0$ signal very different physics on the two sides (see Fig.~\ref{reflectionfig}).
}
\end{figure}

The dynamics are characterized by the weights in the bunched and antibunched waves, which can be extracted from the long-time distribution. In particular, since antibunched waves contain only one particle on each side, the weight in the bunched waves moving left (right) approximately equals the probability of finding both particles in the bosonic (fermionic) side, $\smash{P^{b(f)}_{\sbullet[0.7]\sbullet[0.7]}}$. These are also related to the imbalance $\smash{\mathcal{I} := (n^b-n^f)/(n^b+n^f) = P^b_{\sbullet[0.7]\sbullet[0.7]} - P^f_{\sbullet[0.7]\sbullet[0.7]}}$, where $n^{b(f)}$ is the average number of particles on the boson (pseudofermion) side. As the initial positions cross the boundary, the physics changes drastically, producing sharp variations in $\smash{P^{b(f)}_{\sbullet[0.7]\sbullet[0.7]}}$ and $\mathcal{I}$, as shown in Fig.~\ref{metricsfig}. The asymmetry falls if the release sites are far inside the boson region since the bunching is not perfect and the particles have more time to delocalize. However, we do not see this behavior on the pseudofermion side where the bunched motion is strongly bound. Similarly, $\mathcal{I}$ falls off with the initial separation of the particles (see Supplement \cite{supplement}). Note that $\smash{P^{b(f)}_{\sbullet[0.7]\sbullet[0.7]}}$ and $\mathcal{I}$ are directly measurable in a quantum gas microscope \cite{Preiss2015}.

{\it Statistical well.---}The reflections sketched in Figs.~\ref{reflectionfig}B and \ref{reflectionfig}E lead to striking dynamics when multiple domain walls coexist. Figure~\ref{wellfig}A shows a $\pi$-0-$\pi$ interface, where a bosonic region is sandwiched between two pseudofermion regions, forming a ``statistical well.'' Here, upon release in the middle, the particles repeatedly bounce back-and-forth as bunched waves, as per Fig.~\ref{reflectionfig}B. At each bounce, nearly half the incoming flux leaks out into antibunched motion, giving rise to multiple correlated single-particle waves. The reflections are more prominent for a narrow boson region which reduces delocalization. On the other hand, for a 0-$\pi$-0 interface, shown in Fig.~\ref{wellfig}B, the loss and delocalization are both suppressed (as in Fig.~\ref{reflectionfig}E), but reflections occur on a longer timescale set by $v_{\text{slow}}$. In both cases, the width of the surrounding region is irrelevant as the bunched waves are confined inside the well; this is confirmed by numerics (see Supplement \cite{supplement}).

\begin{figure}
\centering
\includegraphics[width=\columnwidth]{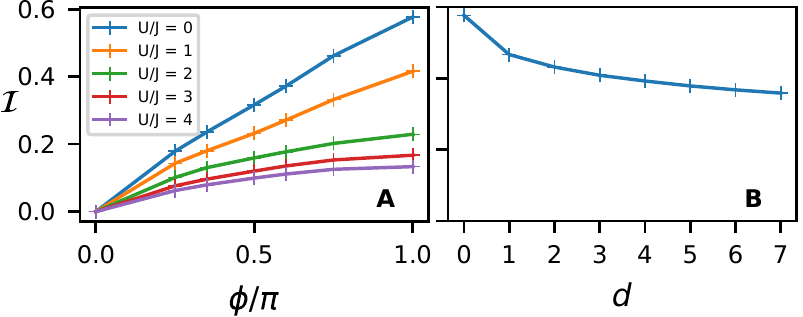}
\caption{\label{expfig}(A) Long-time number imbalance between two sides of a 0-$\phi$ boundary, $\theta_j = 0$ for $j \leq 0$ and $\phi$ for $j>0$, with on-site interactions $U$, after two anyons are released from $j=0,1$. (B) Imbalance for a boundary of finite width $d$, over which $\theta_j$ varies from 0 to $\pi$, for the same initial state and $U=0$.
}
\end{figure}

{\it Experimental considerations.---}So far, we have considered zero on-site interactions and maximum phase jump across a sharp interface. In Fig.~\ref{expfig}, we show these conditions are by no means necessary for observing the physics. Figure~\ref{expfig}A shows the imbalance $\mathcal{I}$ between two sides of a sharp 0-$\phi$ interface with $U\neq 0$, when the particles are released at the boundary (as in Fig.~\ref{asymfig}). As expected, $\mathcal{I}$ falls monotonically as $\phi$ is decreased or $U/J$ is increased. However, the change is gradual and $\mathcal{I}$ remains large even for $U/J \sim 1$. In Fig.~\ref{expfig}B, we consider an interface of finite width $d$, where $\theta_j$ varies linearly from 0 to $\pi$ over $d$ sites. Here, $\mathcal{I}$ decreases slowly and reaches a plateau at large $d$, so the physics is also not sensitive to $d$.

As we stated earlier, of the several protocols for engineering Anyon-Hubbard models \cite{Keilmann2011, Greschner2015, Straeter2016, Cardarelli2016, Greschner2018}, the one in Ref.~\cite{Greschner2015} is most suited for our purpose. Here, one uses a set of Raman lasers to control the tunneling of atoms with two internal states on a tilted optical lattice, such that $U$ is fully tunable by a detuning. The exchange statistics is set by the relative phase of one of the lasers, which can be varied spatially to form a domain wall.

{\it Summary and outlook.---}We have investigated a novel scenario where multiple anyonic regions are separated by domain walls in the same physical system, and particles change their statistics by hopping across a wall. One can engineer this setting via occupation- and site-dependent hopping in realistic atomic setups. We have studied two-body walks in the vicinity of such a wall and showed that the dynamics are marked by a characteristic reflection of bunched waves at the interface that is strongly asymmetric and sensitive to initial conditions, leading to striking phenomena. These reflections leave experimentally measurable signatures in the long-time distribution.

A distinguishing feature of such a statistical interface is that it is, by definition, transparent to single particles and affects the physics only by exchanging multiple particles arriving simultaneously. Thus, our findings strongly encourage future studies of this intrinsically many-body operation. For example, Refs.~\cite{Keilmann2011, Greschner2015, ArcilaForero2016, Lange2017, Cardarelli2016} have found insulating and superfluid phases of anyons as a function of their statistics, which can be combined to explore correlated transport through statistical junctions. Our work also highlights new phenomena that open up by density-dependent gauge fields and motivate further experimental developments at this exciting frontier \cite{Meinert2016, Clark2018, Goerg2019, Schweizer2019, Banuls2020}.

%\begin{acknowledgments}
We thank Nigel Cooper for valuable feedback. This work was supported by the Engineering and Physical Sciences Research Council Grant No. EP/P009565/1.
%\end{acknowledgments}

%\bibliography{references}

\begingroup
\renewcommand{\addcontentsline}[3]{}% Remove functionality of \addcontentsline
\renewcommand{\section}[2]{}% Remove functionality of \section

\endgroup

\onecolumngrid
\clearpage

\begin{center}
\textbf{\large Supplemental Material: ``Quantum walk of two anyons across a statistical boundary''}
\end{center}

    \renewcommand{\thefigure}{S\arabic{figure}}
    \renewcommand{\theHfigure}{S\arabic{figure}}
    \renewcommand{\theequation}{S\arabic{equation}}
    \renewcommand{\thetable}{S\arabic{table}}
    \renewcommand{\thesection}{S\Roman{section}}
    \renewcommand{\bibnumfmt}[1]{[S#1]}
    \renewcommand{\citenumfont}[1]{S#1}
    
    \setcounter{equation}{0}
    \setcounter{figure}{0}
    \setcounter{table}{0}
    \setcounter{page}{1}
    \setcounter{section}{0}
    \setcounter{secnumdepth}{3}
    \makeatletter
    
    \renewcommand{\baselinestretch}{1}\normalsize
    \tableofcontents
    \renewcommand{\baselinestretch}{1.25}\normalsize
    \normalfont
    
\section{Characterization of the slow bunched wave}\label{sec:slow_bunched}
\begin{figure}[H]
    \centering \includegraphics[width = 0.95\columnwidth]{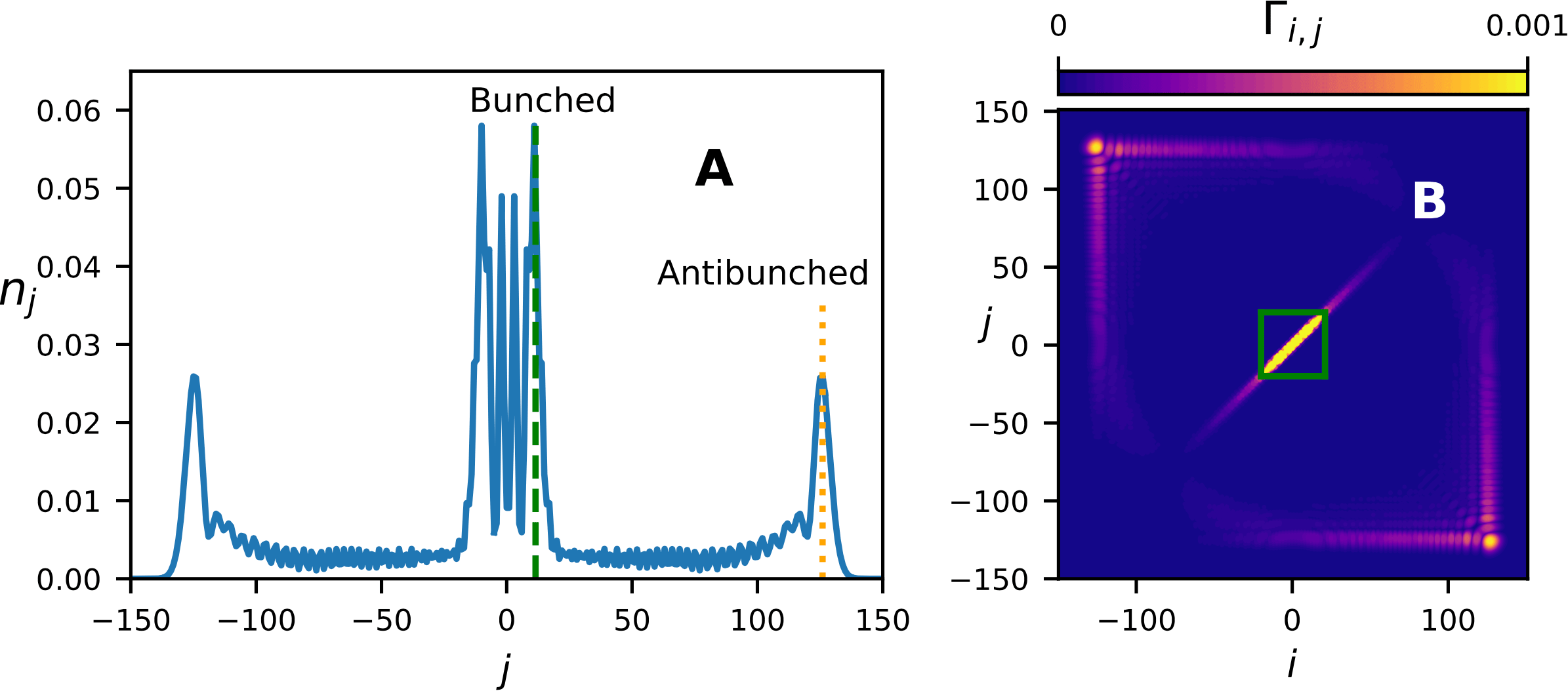}
    \caption{\label{fig:s1}Superposition of fast antibunched and slow bunched propagation with exchange phase $\theta_j = \pi$ at $Jt = 65$ after two anyons with $U=0$ are released from sites $j=0,1$. (A) Density profile showing a slow bunched mode, with sharp leading edges (dashed line), and a fast antibunched mode, peaked at $j=\pm 2J t$ (dotted line). (B) Density-density correlations showing a clear separation between the antibunched wavefront and strongly bound bunched waves within the green square.}
\end{figure}

\begin{figure}[H]
    \centering
    \includegraphics[width =\columnwidth]{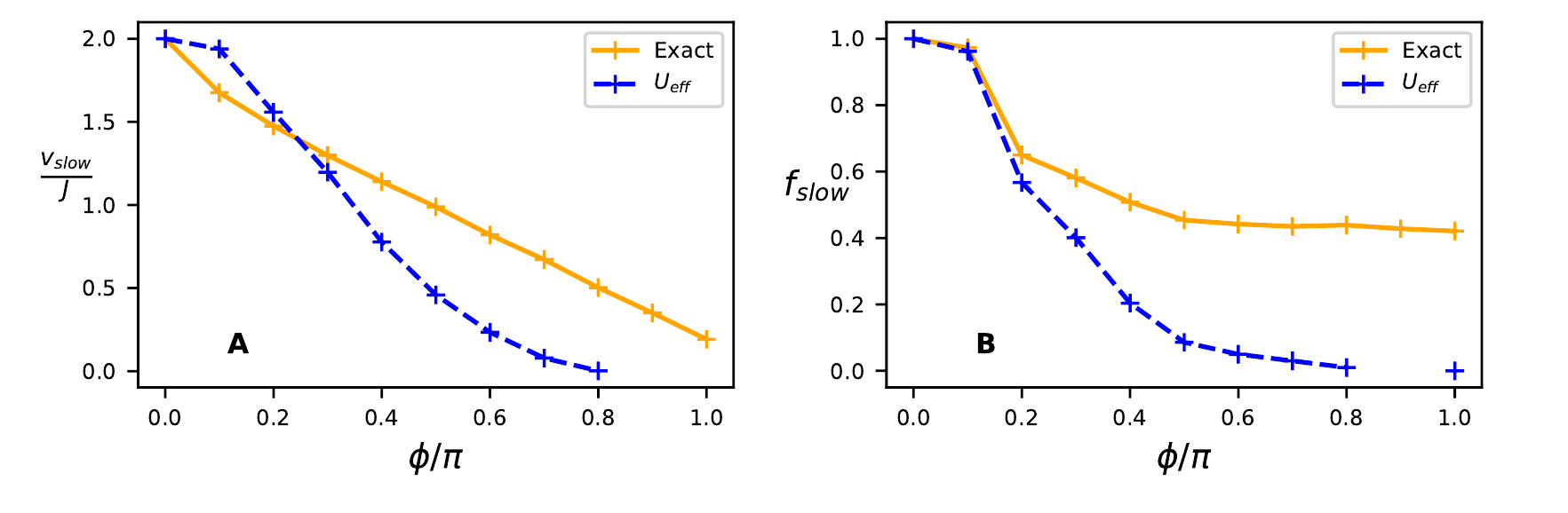}
    \caption{\label{fig:s2}(A) Propagation speed and (B) relative weight of the slow bunched waves as a function of $\theta_j=\phi$ for the exact model (solid lines) and using the effective repulsion in Eq.~\eqref{eq:Ueff} (dashed lines). For the latter, the bunched waves are indistinguishable from background at $\phi>0.8\pi$. Note the strong mismatch between the two curves at large $\phi$. In particular, at $\phi=\pi$, the effective repulsion describes hard-core bosons or free fermions, so double occupancy is prohibited and bunching is not supported. }
\end{figure}

In this section we investigate the effect of the exchange statistics on the speed and relative weight of the slow bunched waves discussed in Fig.~\ref{uniformfig} of the main article. As before, we consider anyons on a 1D lattice with exchange phase $\theta_j$ that map onto bosons with occupation- and site- dependent Peierls phase, described by the Hamiltonian
\begin{equation}
\hat{H} = 
-J \sum_j \big(\hat{b}_j^{\dagger} \hat{b}_{j+1} e^{{\rm i} \theta_j \hat{n}_j} + \text{H.c.}\big) 
+ \frac{U}{2} \sum_j \hat{n}_j (\hat{n}_j -1) \push,
\label{eq:hamilboson}
\end{equation}
where $J$ is the tunneling, $U$ is the interaction, $\smash{\hat{b}_j^{\dagger}}$ and $\smash{\hat{b}_j}$ are boson creation and annihilation operators, and $\smash{\hat{n}_j := \hat{b}_j^{\dagger} \hat{b}_j}$. We consider anyons with $U=0$ in the uniform case $\theta_j = \phi$, where nonzero $\phi$ leads to a separation of timescales in the evolution: As shown in Fig.~\ref{fig:s1}, when two particles are released from neighboring sites, the dynamics split into a fast antibunched wave, where the two anyons travel in opposite directions at speed $2J$, and a slow bunched wave, where the anyons are strongly bound and propagate more slowly.
We use the leading edge of the bunched wave in the density profile (Fig.~\ref{fig:s1}A) to characterize its speed as a function of the statistical phase $\phi$. As shown in Fig. \ref{fig:s2}A, this speed falls off linearly from $v_{\text{slow}} = 2J$ at $\phi=0$ to $v_{\text{slow}} \approx J/5$ at $\phi=\pi$. We calculate the relative weight in the bunched mode from the two-body correlations $\Gamma_{i,j} := \langle\hat{n}_i \hat{n}_j \rangle - \delta_{ij} \langle\hat{n}_j\rangle$ as $f_{\text{slow}} = \sum_{\Box} \Gamma_{i,j}/2$, where $\Box$ encloses the bunched waves (see Fig.~\ref{fig:s1}B; note that $\sum_{i,j} \Gamma_{i,j}=2)$. Figure~\ref{fig:s2}B shows that $f_{\text{slow}}$ decreases monotonically with $\phi$, saturating around $0.5$. For comparison, we also plot the speed and weight using an effective repulsion instead of nonzero $\phi$ \cite{sGreschner2015}, as discussed in the main text and elaborated in the next section.

\section{Effective repulsion}\label{sec:effective_coupling}
Here we provide quantitative comparisons between the exact dynamics and that generated by an effective interaction derived from the scattering length. As detailed in Ref.~\cite{sGreschner2015}, the low-energy collisions between two anyons with exchange phase $\phi$ is characterized by the scattering length (in units of the lattice spacing)
\begin{equation}
    a_s = \frac{-\left(1+ \cos \phi \right)}{4\left(1-\cos \phi\right) + 2U/J} \;, \label{eq:scatt_length}
\end{equation}
which can be interpreted as originating from an effective interaction strength% as a function of interaction strength and exchange phase is
\begin{equation}
    U_{\text{eff}}  = \frac{4J \left(1-\cos \phi \right) +2U}{\left(1+\cos \phi\right)} \push. \label{eq:effective_repulsion}
\end{equation}
Note that $U_{\text{eff}} \vert_{\phi\to 0} = U$. In the limit $U \to 0$, we obtain an effective repulsion as a function of the exchange phase,
\begin{equation}
U_{\text{eff}} = 4J \tan^2(\phi/2)\;,
\label{eq:Ueff}
\end{equation}
which can lead to slow-moving repulsively bound pairs \cite{sPreiss2015} through the effective Hamiltonian
\begin{equation}
    H_{\text{eff}} = -J \sum_j \big(\hat{b}_j^{\dagger} \hat{b}_{j+1} + \text{H.c.}\big) + \frac{U_{\text{eff}}}{2} \sum_j \hat{n}_j (\hat{n}_j -1) \;. \label{eq:h_eff}
\end{equation}
However, the comparisons in Fig.~\ref{fig:s2} show this effective repulsion does not capture the full dynamics when the anyons are released from adjacent sites. In particular, for angles close to $\pi$, $U_{\text{eff}}$ diverges and does not support any bunched propagation, which is a crucial feature of the model in Eq.~\eqref{eq:hamilboson}. This breakdown is not surprising since the scattering length in Eq.~\eqref{eq:scatt_length} becomes less than the lattice spacing (in magnitude) for $\phi \gtrsim 1$, changing its interpretation \cite{sKolezhuk2012}.
\begin{figure}[H]
    \centering
    \includegraphics[width = 0.9\columnwidth]{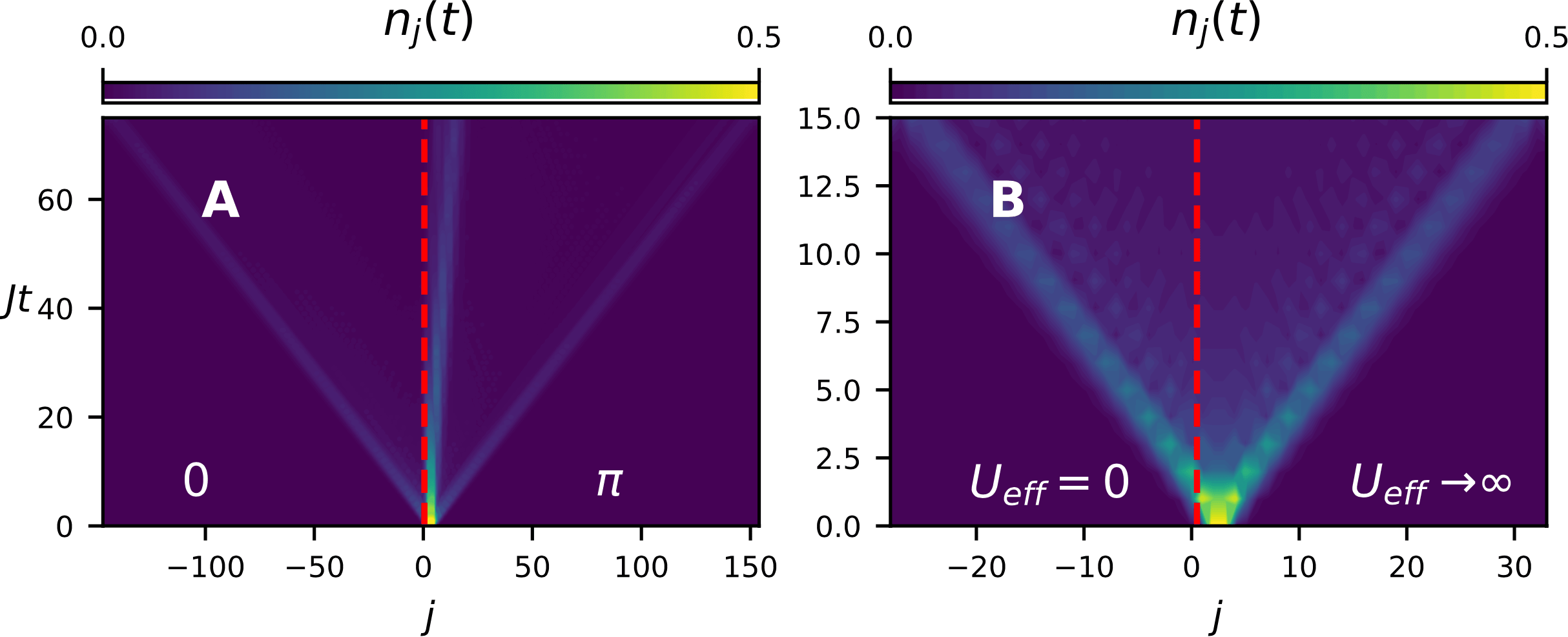}
    \caption{\label{fig:s3}Evolution of the density profile in the presence of a $0$-$\pi$ statistical interface ($\theta_j = \pi$ for $j > 0$), following the release of two particles from sites $j = 3,4$ with $U=0$. (A) Exact model [Eq.~\eqref{eq:hamilboson}], showing a complete reflection of the slow bunched waves. (B) Using effective repulsion [Eq.~\eqref{eq:Ueff}], showing a fully antibunched symmetric walk that is not affected by the interface.}
\end{figure}
\par The above discrepancy leads to very different dynamics in the vicinity of a $0$-$\pi$ statistical interface when the particles are released from the pseudofermionic ($\theta_j = \pi$) side, as shown in Fig.~\ref{fig:s3}. For the exact model, nearly half the initial weight goes into the slow bunched mode, which is completely reflected at the boundary, as explained in the main text. On the other hand, with the effective repulsion, the particles are fully antibunched like free fermions: they travel separately in opposite directions and are not affected by the statistical boundary.%, yielding a symmetric walk.
\begin{figure}[H]
    \centering
    \includegraphics[width=0.9\columnwidth]{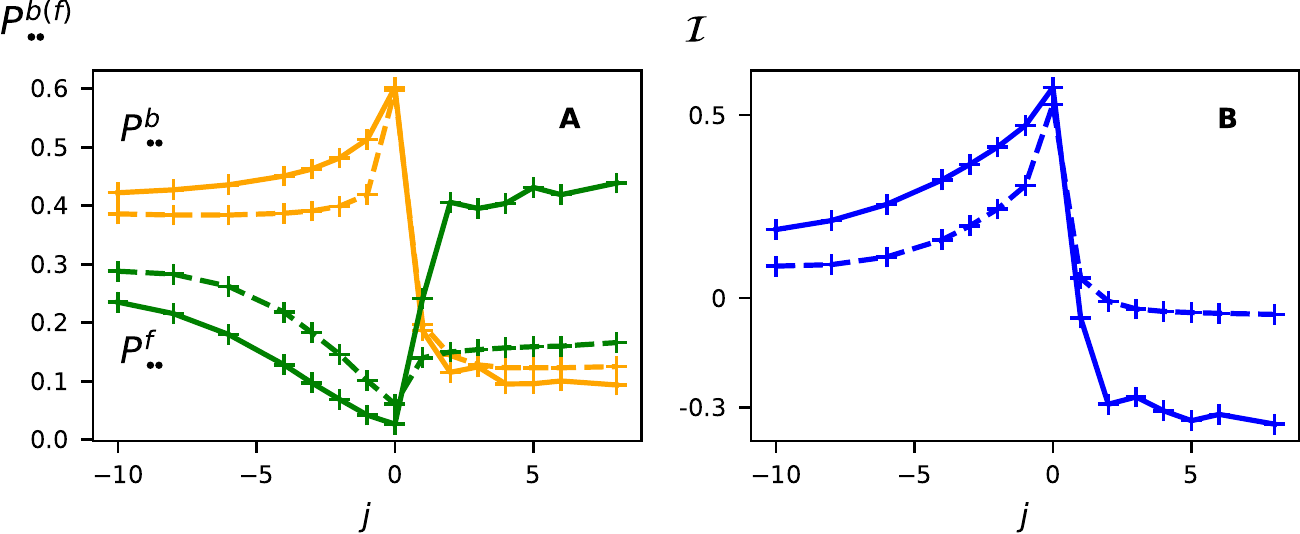}
    \caption{\label{fig:s4}Long-time asymmetries ($Jt=100$) after the anyons are released from sites $j,j+1$ with the $0$-$\pi$ interface in Fig.~\ref{fig:s3} for the exact model (solid) and with corresponding effective interactions (dashed). (A) Probability of finding both particles in the left (orange) and right (green) halves. (B) Relative number imbalance between the two halves. The effective repulsion captures the qualitative features for bosonic initial states ($j\leq 0$) but produces little asymmetry for pseudofermionic initial states.}
\end{figure}
\par In Fig.~\ref{fig:s4}, we show how the long-time asymmetries in the distribution are modified in the effective-repulsion picture for different initial positions $(j,j+1)$ around a $0$-$\pi$ interface. As in the main text, we consider the metrics $\smash{P^{b(f)}_{\sbullet[0.7]\sbullet[0.7]}}$ and $\mathcal{I}$, where $\smash{P^{b(f)}_{\sbullet[0.7]\sbullet[0.7]}}$ is the probability of finding both particles in the bosonic (fermionic) side and $\smash{\mathcal{I} := (n^b-n^f)/(n^b+n^f)}$ is the relative number imbalance between the two sides. Note the exact dynamics always produce stronger asymmetries. The agreement between the two is better when the particles are released from the bosonic side ($j\leq 0$) as the effective repulsion can approximately reproduce the reflection of the bunched waves. For $j>0$, these are absent as in Fig.~\ref{fig:s3}B, hence the imbalance $\mathcal{I}$ is very close to zero for the effective repulsion. Note the same-side probabilities $\smash{P^{b(f)}_{\sbullet[0.7]\sbullet[0.7]}}$ are still nonzero since the antibunching is not perfect and the particles are delocalized.

\section{Symmetric initial states with larger separation}\label{sec:dropoff_initial_distance}
\begin{figure}[H]
    \centering \includegraphics[width = \textwidth]{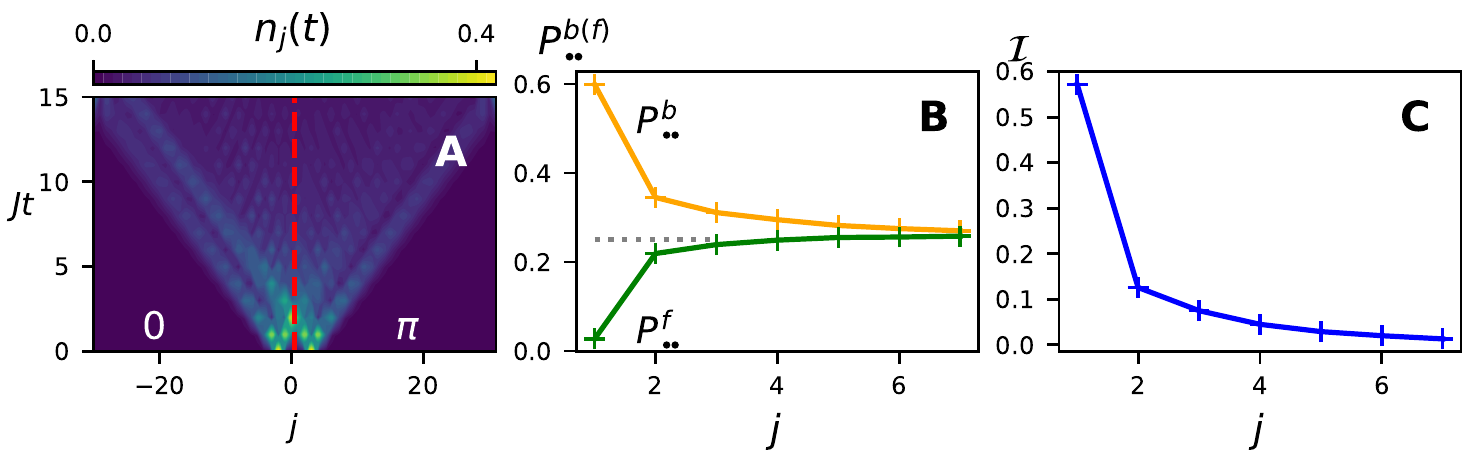}
    \caption{\label{fig:s5}Time evolution and long-time asymmetries after the anyons are released from sites $-j+1, j$ on opposite sides of a $0$-$\pi$ interface, with $U=0$. (A) Density profile, showing how the interface gives rise to an asymmetry by sending bunched waves preferentially into the bosonic region. (B) Same-side probabilities $\smash{P^{b(f)}_{\sbullet[0.7]\sbullet[0.7]}}$ and (C) relative number imbalance $\mathcal{I}$, showing how the asymmetry falls off with larger initial separation. For $j\to\infty$, the particles move independently, so $\smash{P^{b(f)}_{\sbullet[0.7]\sbullet[0.7]}} \to 1/4$ (dotted line).}
\end{figure}

Here we show how the effect of a statistical boundary is reduced for larger initial distance between the two particles. As before, we focus on a $0$-$\pi$ interface at $j=0$. Since it alters the dynamics only by swapping the particles, we consider symmetric initial states, where two particles are released from sites $-j+1,j$, to ensure they arrive at the boundary simultaneously from the left and the right. Figure~\ref{fig:s5}A shows that the interface produces an asymmetry by sending some of this incident weight as bunched waves toward the bosonic region. As the initial separation is increased, the asymmetry falls off as shown in Figs.~\ref{fig:s5}B and \ref{fig:s5}C. This is because the particles have more time to delocalize before arriving at the boundary, so less weight arrives as bunched. For large separation, the particles evolve independently and the exchange phase is redundant, so $\smash{P^{b(f)}_{\sbullet[0.7]\sbullet[0.7]}} \pull \to (1/2)^2$. Note the separation is an odd number for the symmetric states, but we find a similar behavior when the particles are initially separated by an even number of sites.

\section{Reflections off a statistical region of finite width}\label{sec:width}
\begin{figure}[H]
    \centering \includegraphics[width = 0.9\textwidth]{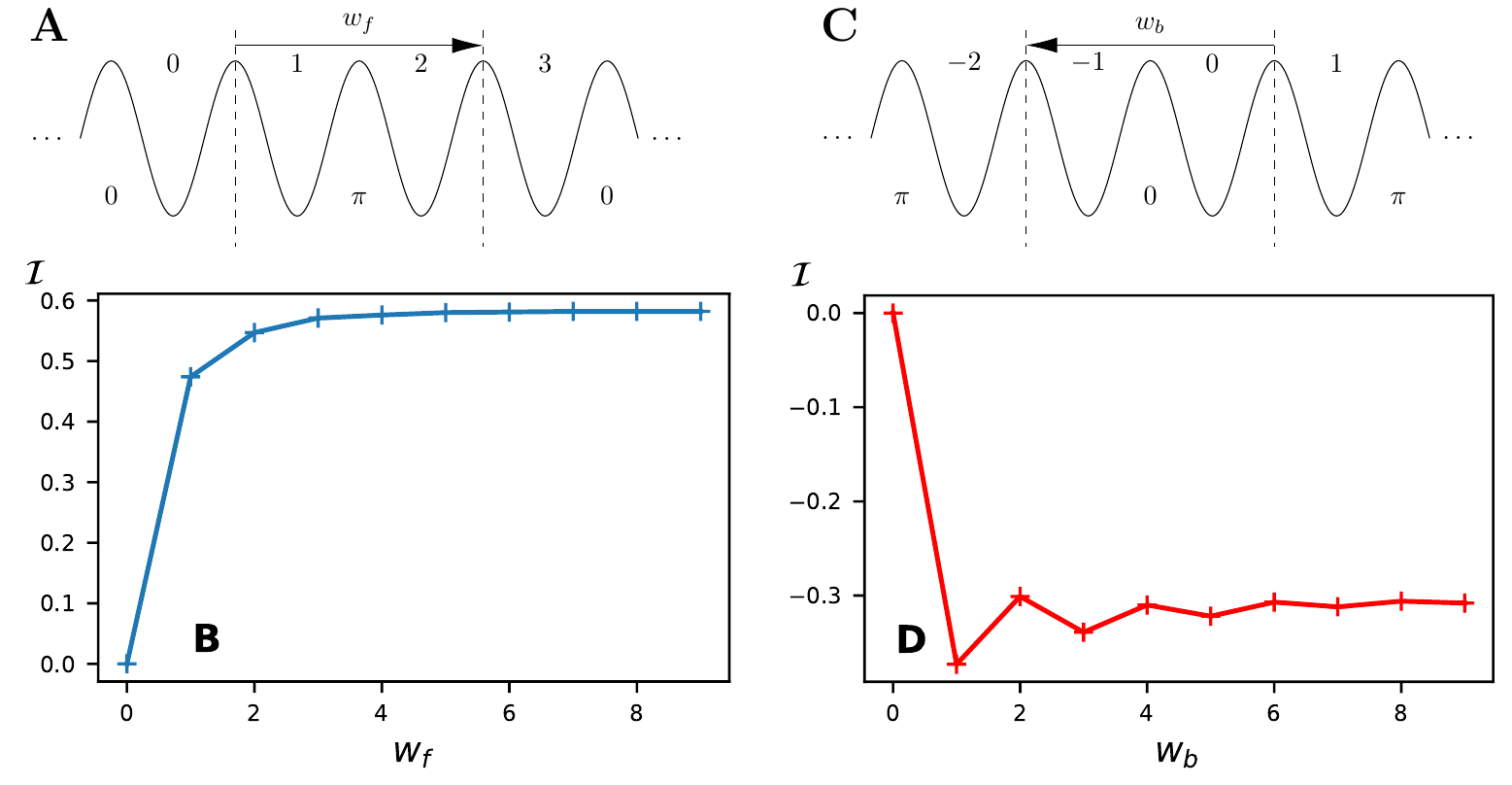}
    \caption{\label{fig:s6}Insensitivity of the dynamics of two anyons with $U=0$ released from one side of a statistical junction to the width of the middle region. (A) 0-$\pi$-0 interface; anyons released from $j=0,1$, and (C) $\pi$-0-$\pi$ interface; anyons released from $j=2,3$. (B) and (D) show the corresponding number imbalance at long times ($Jt=100$) between the regions $j \leq 0$ and $j>0$. The imbalance saturates to a nonzero value as soon as the middle region spans two or more sites.}
\end{figure}

Consider a junction of statistical regions $\alpha$-$\beta$-$\gamma$ with exchange phases (I) $\phi_{\alpha} = 0$, $\phi_{\beta} = \pi$, $\phi_{\gamma} \neq \pi$ or (II) $\phi_{\alpha} = \pi$, $\phi_{\beta} = 0$, $\phi_{\gamma} \neq 0$, and suppose the particles are released in region $\alpha$. Here we show the dynamics are insensitive to the width of region $\beta$. This is expected since no bunched waves are transmitted through the $\alpha$-$\beta$ interface, as sketched in Fig.~\ref{reflectionfig} of the main text, which makes the $\beta$-$\gamma$ interface redundant. This is numerically confirmed in Fig.~\ref{fig:s6} which shows that the long-time asymmetry quickly saturates as a function of the width of region $\beta$ for both (I) with $\phi_{\gamma} = 0$ and (II) with $\phi_{\gamma} = \pi$.

\begingroup
\renewcommand{\addcontentsline}[3]{}% Remove functionality of \addcontentsline
\renewcommand{\section}[2]{}% Remove functionality of \section

\endgroup

\end{document}